\documentclass[%
 preprint,
 amsmath,amssymb,
 aps, physrev,
]{revtex4-2}

\usepackage{graphicx}
\usepackage{dcolumn}
\usepackage{bm}
\usepackage{color}
\usepackage{bm}
\usepackage{graphicx}
\usepackage{etoolbox}
\usepackage{calc}

\begin{document}


\title{\textbf{Motorized Chromosome Models of Mitosis} 
}%

\author{Zhiyu Cao}
\affiliation{%
 Center for Theoretical Biological Physics, Rice University, Houston, TX 77005
}%

\author{Chaoqun Du and Zhonghuai Hou}
\affiliation{%
 Department of Chemical Physics, University of Science and Technology of China, Hefei, Anhui 230026
}%

\author{Peter G. Wolynes}
\email{Contact author: pwolynes@rice.edu}
\affiliation{
Center for Theoretical Biological Physics, Rice University, Houston, TX 77005
}%
\affiliation{
Department of Chemistry, Rice University, Houston, TX 77005
}%
\affiliation{
Department of Physics, Rice University, Houston, TX 77005
}%


\date{\today}



\begin{abstract}
During mitosis, near-spherical chromosomes reconfigure into rod-like structures to ensure their accurate segregation to daughter cells. We explore here, the interplay between the nonequilibrium activity of molecular motors in determining the chromosomal organization in mitosis and its characteristic symmetry-breaking events. We present a hybrid motorized chromosome model that highlights the distinct roles of condensin I and II in shaping mitotic chromosomes. Guided by experimental observations, the simulations suggest that condensin II facilitates large-scale scaffold formation, while condensin I is paramount in local helical loop arrangement. Together, these two distinct grappling motors establish the hierarchical helical structure characteristic of mitotic chromosomes, which exhibit striking local and, sometimes global, chirality and contribute to the robust mechanical properties of mitotic chromosomes. Accompanying the emergence of rigidity, the model provides mechanisms of forming defects, including perversions and entanglements, and shows how these may be partially resolved through condensin activity and topoisomerase action. This framework bridges coarse-grained energy landscape models of chromosome dynamics and non-equilibrium molecular dynamics, advancing the understanding of chromosome organization during cell division and beyond. 
\end{abstract}

\maketitle

\section{Introduction}
When cells divide, their chromosomes undergo significant conformational changes. During mitosis, nearly spherical chromosomes condense into the characteristic cylindrical shapes visible under a microscope before forming a pair of sister chromatids \cite{marko1995stretching}. Understanding how these cylinders form is a fundamental question. From the perspective of polymer theory, it is surprising that compaction results in cylindrical rather than spherical shapes. As self-attracting polymers usually collapse into spherical globules \cite{de1979scaling}, clearly forming cylinders involves breaking rotational symmetry. Furthermore, recent experimental observations have shown that during mitosis, as the features corresponding to topologically associated domains (TADs) \cite{dixon2012topological,nora2012spatial,olivares2016capturing} and A/B compartments \cite{lieberman2009comprehensive} disappear, a secondary diagonal band emerges \cite{gibcus2018pathway}. These observations suggest chromosomes are tightly arranged arrays of chromatin loops \cite{marsden1979metaphase,earnshaw1983architecture,naumova2013organization,nagano2017cell}.

The transition from interphase to mitosis involves numerous proteins. Over 4000 proteins, categorized into 28 functional groups, have been found to associate with chromosomes during mitosis \cite{ohta2010protein}. Among the implicated proteins there are histones, condensin and cohesin complexes, as well as DNA topoisomerase II alpha \cite{paulson2021mitotic,meijering2022nonlinear}. Gibcus et al. have suggested that the condensation may be thought of as being meditated by the loop extrusion processes, primarily employing two types of condensins \cite{gibcus2018pathway}. They suggest condensin II binds during prophase to form an axial scaffold and condensin I then regulates the size and nested arrangement of chromatin loops within the helices. Other experimental observations \cite{green2012contrasting} and molecular dynamics simulation results \cite{forte2024bridging} also come to a similar viewpoint. The information theoretical energy landscape derived from Hi-C data using the maximum entropy method, while being mechanistically agnostic, successfully reproduces the known structural changes and patterns of symmetry breaking \cite{zhang2016shape,contessoto2024energy}. Since condensins are ATP-dependent, however, condensation is an active process, not strictly at thermodynamic equilibrium \cite{alipour2012self,swedlow2003making,goloborodko2016chromosome,goloborodko2016compaction,michieletto2016polymer,vian2018energetics,goychuk2023polymer,goychuk2024delayed,bonato2021three,miermans2020lattice,takaki2021theory}. In this paper, we explore how active mechanisms can lead to the symmetry-breaking events of mitosis.

Here, we describe the structural ensembles of mitotic chromosomes induced by ``grappling motors."  ``Grappling motors" provide a caricature of the loop extrusion processes driven by SMC complexes \cite{cao2024motorized,wang2012active}.  Guided by the experimentally inferred scenarios, we explicitly introduce two types of motors with distinct ranges of action corresponding to condensin I and condensin II \cite{gibcus2018pathway,hoencamp20213d}. When both motors are active, chromosomes take on their characteristic rod-like cylindrical shapes. The resulting structural ensemble displays hierarchical helices that can be interpreted as continuously arranged loops, with an additional global helical structure that would naturally lead to broken chiral symmetry. The two motors play distinct roles in mitotic chromosome organization. Condensin II enhances bending rigidity and compaction density, thereby aiding the chromosome in maintaining its structural integrity when pulled or stretched. Condensin I, assisted by topoisomerase in the present model, partially untangles the chromosome, increasing chromatin accessibility and further facilitating separation. The structural ensembles are not perfect. The simulations reveal that there are several defects in the resulting mitotic structures. We focus particularly on the nature of perversions which are defects of chirality and also on the knots remaining in the structure. The heterogeneity of motorization may play a big role in this defect structure. By comparing our results with experimental observations, we propose that chromosomes first become compact during prophase, forming a super-entangled but still cylindrical structure upon condensin II binding. This structure subsequently evolves during prometaphase, when condensin I and topoisomerases help establish the more organized mitotic structure with global chirality.    

\section{Background and model}
Condensin I and II, as parts of the structural maintenance of chromosomes (SMC) complexes family, act by extruding DNA loops, which help organize and segregate chromosomes during mitosis \cite{fudenberg2016formation,ganji2018real,davidson2019dna,kim2019human,wang2017bacillus,davidson2021genome,banigan2020loop,wang2017bacillus,ryu2022condensin,ryu2021bridging,brandao2021dna,dekker2024chromosome,polovnikov2018effective,polovnikov2023topological,belan2024footprints,starkov2024effect,fujishiro2024three}. These complexes consume chemical energy \cite{vian2018energetics,takaki2021theory,zwicker2022intertwined,marko2019dna}, driving non-equilibrium reactions and thus exerting mechanical forces \cite{fudenberg2016formation,rao20143d,sanborn2015chromatin,needleman2017active,van2019role,struhl2013determinants,kschonsak2015shaping,hirano2016condensin}. One specific motorized mechanism may be summarized as the ``swing and clamp" model \cite{bauer2021cohesin}. In this model, the ATPase head of the motor first clamps a DNA segment, after which the hinge domain extends and swings to grasp a nearby segment. The hinge then brings the new segment to the head, merging it into the DNA loop. Experiments and simulations strongly suggest that the coiled-coil part of the SMC complex's structure is a key here. It acts not merely as a linker but actively participates in loop extrusion by twisting and bending, effectively remodeling DNA topology \cite{eeftens2016condensin,kulemzina2016reversible,hons2016topology,burmann2017tuned,krepel2018deciphering,yamauchi2024smc,krepel2020braiding,janissen2024all,takaki2021theory}. Similar mechanisms have also described as the ``reel and seal" and ``hold and feed" models \cite{dekker2023molecular,shaltiel2022hold}. At the coarse-grained level, these mechanisms are equivalent and we characterize them as being ``grappling motor" mechanisms \cite{cao2024motorized}.

\begin{figure}[htbp]
\renewcommand*{\thefigure}{\arabic{figure}}
\centering
\includegraphics[width=0.9\columnwidth]{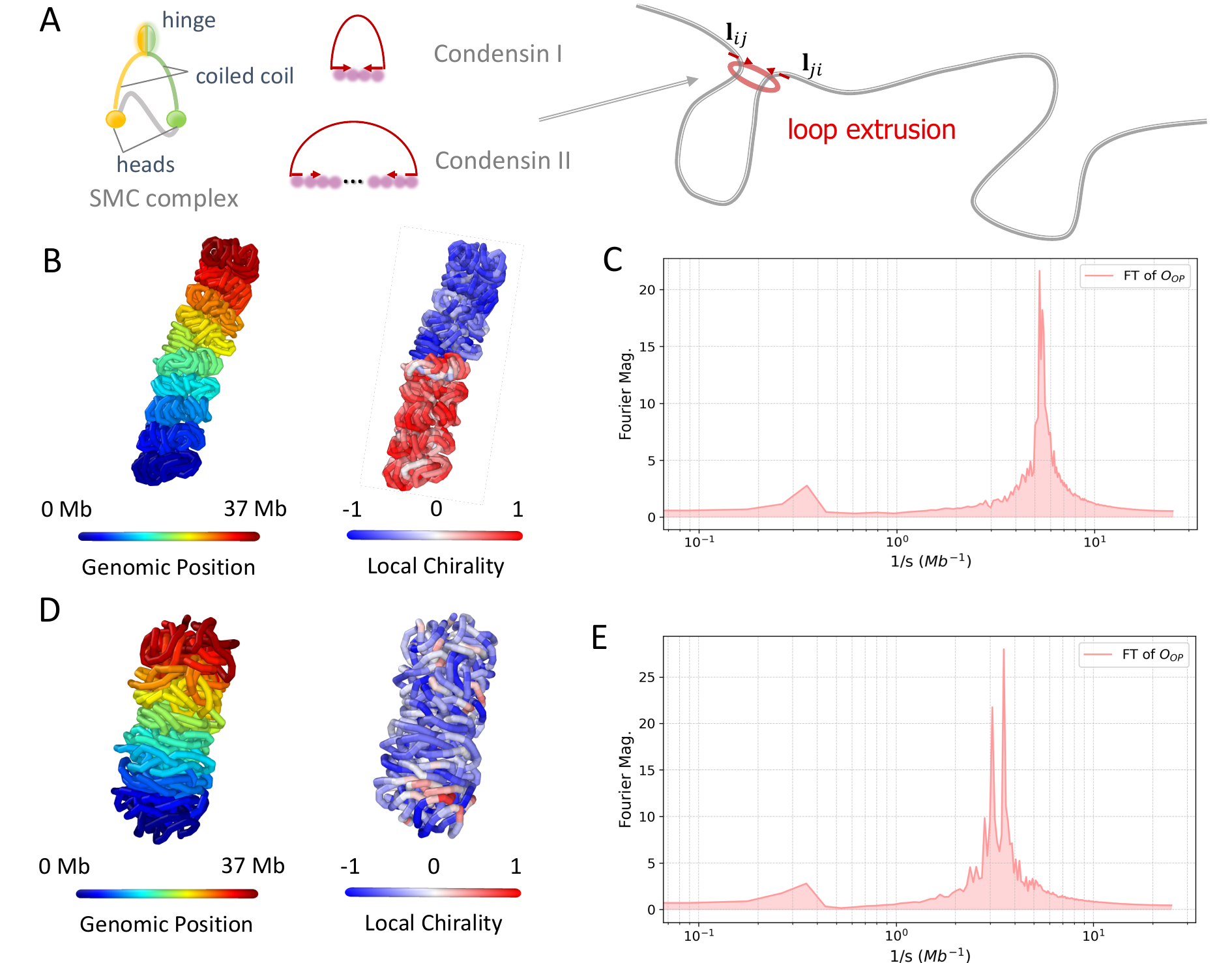}
\caption{\textbf{The mitotic chromosome folds under the combined action of condensin I and II.} (A) The motorized chain model of chromosomes. Chromosome dynamics involve diffusion under a homopolymer potential and non-equilibrium motorization. Structural Maintenance of Chromosomes (SMC) complexes actively extrude loops by consuming ATP, interacting with DNA fragments. We model the loop extrusion process as a grappling motion, where fragments separated by a certain distance can be brought together. During mitosis, condensin I and condensin II cooperate to drive the structural formation of chromosomes. Condensin II binds during prophase, mediating long-range grappling motions to form an axial scaffold. Condensin I binds during prometaphase, regulating chromatin loops' size and nested arrangement within local helices. (B), (D) The representative structures under the combined action of condensin I and II. (B) The purely sequence-distance-dependent model ($\kappa_I=5$ and $\kappa_{II}=2.5$); (D) The spatial-distance-dependent model ($\kappa_I=\kappa_{II}=0.5$). The color bars indicate genomic sequences (left panel) and local chirality (right panel) variations along the genomic sequence. (C), (E) The Fourier transform of the representative structure's orientational order parameter along the genomic position for the purely sequence-distance-dependent model (C) and the spatial-distance-dependent model (E).}
\label{fig:1} 
\end{figure}

In this paper, we study how condensin I and II mediate the formation of mitotic chromosomes through a hybrid coarse-grained motorized chromosome model. A schematic representation is shown in Fig.\ref{fig:1}. The chromosome is described by a coarse-grained bead-spring homopolymer with bead size $\sigma$, which is acted upon by stochastic motorization. The motion of the chromosomal backbone is described by an overdamped Langevin equation $\dot{\mathbf{r}}_{i}=\beta D(-\mathbf{\nabla}_i U_{HP})+\bm{\eta}_i+\mathbf{v}_i^{m}$, where the last term describes the crucial addition of the motorized grappling events. Here, $\mathbf{r}_i$ is the position of the $i$-th bead. The details of the homopolymer potential used here are outlined in \textcolor{blue}{Supplementary Materials (SMs) and Methods}. It is worth emphasizing that our homopolymer potential includes a soft-core component $U_{sc}$, which while providing excluded volume effects, allows chains to pass through each other at only a finite energy cost. The softness of this repulsion mimics the way topoisomerases can act to allow chains to pass through each other. The random variable $\mathbf{\eta}$ encodes the ordinary Gaussian thermal noise from the solvent with zero average and variance $\langle\eta^{j_1}_{i_1}(t)\eta^{j_2}_{i_2}({t^\prime})\rangle=2D\delta_{i_1i_2}\delta_{j_1j_2}\delta(t-t^{\prime})$,  where $D$ is the diffusion coefficient and $\beta$ is the inverse temperature. The motorized jump term $\mathbf{v}_i^{m}=\sum_q\mathbf{l}_q\delta(t-t_q)$ is a time series of shot-noise-like kicks, which describe the net effect of binding a grappling motor which induces large scale chromosome motions.

The stochastic nature of motorized dynamics encoded in the kicks $\mathbf{v}_i^{m}$ can be described using a Master equation. This equation governs the probability distribution function $\Psi$ of the chain, with its positions in three-dimensional space $\{\mathbf{r}\}$, expressed as $\partial_t\Psi_t(\{\mathbf{r}\})=(\mathcal{L}_{FP}+\mathcal{L}_{NE})\Psi_t(\{\mathbf{r}\})$. The first contribution of the time development comes from the Fokker-Planck operator $\mathcal{L}_{FP}\Psi_t(\{\mathbf{r}\})=-\sum_i\nabla_i\cdot [\beta D(-\nabla_i U)\Psi_t(\{\mathbf{r}\})-D\nabla_i\Psi_t(\{\mathbf{r}\})]$, which describes the passive thermal Brownian dynamics. The other nonequilibrium term describes the motorized displacements of the chain by the motors which are regarded as making discrete jumps of fairly large distance, which can be quantitatively described by the Master equation $\mathcal{L}_{NE}\Psi_t(\mathbf{r})=\int\Pi_i d\mathbf{r}^{\prime}_i K_{\mathbf{r}^{\prime}\to\mathbf{r}}\Psi_t(\mathbf{r}^{\prime})-\Psi_t(\mathbf{r})\int\Pi_i d\mathbf{r}^{\prime}_i K_{\mathbf{r}\to\mathbf{r}^{\prime}}$. The transition probability between different chromosome configurations $K_{\mathbf{r}^{\prime}\to\mathbf{r}}$ encodes how the motors kinematically respond to externally imposed forces. We write $K=\kappa e^{-\vartheta\beta\Delta U}$, where $\kappa$ is the basal grappling rate and $\Delta U$ is the difference between the free energy of the starting configuration and that of the chain configuration that would arise from the motorization displaced hopping event. The parameter $\vartheta$ is the susceptibility of the motor, which can in principle be obtained from the force-extension measurements \cite{dogterom1997measurement,howard2002mechanics}, quantifying the mechanical coupling between the motorized conformational remodeling and the local mechanical forces acting on the motor. When $\vartheta=0$, the motor does not respond at all to imposed forces. We say the motor is fully adamant. When $\vartheta\neq0$, the motor does respond to applied forces by speeding or slowing. Motors with both signs of $\vartheta$ are known. In a more general description, the susceptibilities may take on different values for uphill kicks ($\vartheta_u$) and downhill kicks ($\vartheta_d$).

To depict the loop extrusion events, we model the motorized events as anti-correlated grappling taking place between pairs of nodes $(i,j)$ on the chain. Node $i$ corresponds to the anchor point which becomes bound to the condensin's head, while the node $j$ represents the segment that is eventually grasped by the condensin's hinge. A clamped kick is naturally defined by a pair of displacements along the line of centers $(\mathbf{l}_{ij},\mathbf{l}_{ji})=l(\hat{\mathbf{r}}_{ij},-\hat{\mathbf{r}}_{ij})$, where $l$ is the grappling distance which should be roughly the length of the coiled coils of an SMC complex $L_{SMC}\sim50$nm, and $\hat{\mathbf{r}}_{ij}$ is a unit vector pointing from node $i$ to node $j$, see Fig.\ref{fig:1} A.

The formation of mitotic chromosome structure depends on the simultaneous presence of condensin I and condensin II in the system These two motors appear to act on different scales: condensin I induces the formation of local helical structures, while condensin II is responsible for the formation of the global chiral structure. We ultimately then can write down the complete dynamics, explicitly incorporating the driving of both types of condensin through an operator equation:
\begin{equation}
       \mathcal{L}_{NE}\Psi_t(\mathbf{r})=\frac{1}{2}\sum_\alpha\kappa_\alpha\sum _{i,j}
\iint d\mathbf{r}^{\prime}_id\mathbf{r}^{\prime}_j\mathbf{J}^\alpha_{ij}(\mathbf{r}),
\end{equation}
where $\alpha\in\{I,II\}$ the probability current $\mathbf{J}^\alpha_{ij}(\mathbf{r})$ is
\begin{equation}
\begin{aligned}
\mathbf{J}^\alpha_{ij}(\mathbf{r})&=\delta(\mathbf{r}_i-\mathbf{r}_i^\prime-\mathbf{l}_{ij})\delta(\mathbf{r}_j-\mathbf{r}_j^\prime+\mathbf{l}_{ij}) \mathcal{C}^\alpha(r^\prime_{ij})e^{-\vartheta\beta[U(\mathbf{r}_i,\mathbf{r}_j)
-U(\mathbf{r}_i^\prime,\mathbf{r}_j^\prime)]}\Psi_t(\mathbf{r}_i^{\prime})\\
&-\delta(\mathbf{r}_i-\mathbf{r}_i^\prime+\mathbf{l}_{ij})\delta(\mathbf{r}_j-\mathbf{r}_j^\prime-\mathbf{l}_{ij})\mathcal{C}^\alpha(r_{ij})e^{-\vartheta\beta[U(\mathbf{r}_i^\prime,\mathbf{r}_j^\prime)-U(\mathbf{r}_i,\mathbf{r}_j)]}\Psi_t(\mathbf{r}_i). \label{J}
\end{aligned}
\end{equation} 
We have omitted writing the particles' positions other than those involved in the grappling events in the expression of free energy and configuration to simplify the notation. Here,  $\mathcal{C}^\alpha({r}_{ij})$ is the grappling probability, which describes the rate of occurrence of a loop grappling event involving the $i$-th and $j$-th beads and is generally a function of distance $r_{ij}=|\mathbf{r}_i-\mathbf{r}_j|$. The rate of potential grappling events depends not only on the basal grappling rate but also on the instantaneous chromosome conformation and motor susceptibility. 

Spatially proximal sites are more likely to undergo grappling events. In a fashion like that employed in the ideal chromosome model based on information theoretical energy landscape theory \cite{zhang2016shape,zhang2015topology,di2016transferable}, the spatial-distance-dependent grappling model can be projected onto a roughly equivalent sequence-distance-dependent counterpart by means of a mean-field approximation, where sequence translational invariance is restored \cite{cao2024motorized}. Within this picture, the grappling motor, in a sense, behaves more like an extruder than a grappler. It is important to emphasize that in the purely sequence-distance-dependent model, grappling events still do depend on spatial distance because the probability flux in Eq. \ref{J} is configuration-dependent. Our previous work demonstrated that in the purely sequence-distance-dependent grappling model, the grappling-induced effective diffusion coefficient and effective temperature are configuration-dependent, in contrast to what happens for conventional hydrodynamic interactions that couple bead motion \cite{cao2024motorized}. If motors remain anchored at chromosomal sites for extended periods \cite{walther2018quantitative}, the spatial-distance-dependent grappling model will naturally exhibit processivity \cite{chan2023theory,chan2024activity,tortora2024physical,rahmaninejad2024dynamic}. Successive grappling events then become correlated, introducing a history dependence that implies non-Markovian behavior. Consequently, the purely sequence-distance-dependent model under mean-field approximation will typically exhibit a higher renormalized basal grappling rate and have a larger renormalized grappling distance than that of the spatial-distance-dependent grappling model. We have also previously established a mapping between the sequence-dependent model and the ideal chromosome model theoretically to demonstrate that the former can indeed serve as the origin of the latter \cite{cao2024motorized}, which has already been successfully employed to reconstruct chromosome configurations \cite{zhang2016shape,contessoto2024energy,zhang2015topology,di2016transferable,contessoto2023interphase,di2018anomalous}. Our present analysis indicates that grappling induces effective attraction between loci for both sequence-distance-dependent and spatial-distance-dependent models. The attraction shrinks the chromatin chain, leading to a smaller effective Kuhn length. In the spatial-distance-dependent model, the grappling motors will also induce condensation and further increase the frequency of grappling events between loci, creating a positive feedback loop (a similar mechanism applies to the sequence-distance-dependent model, as effective attraction strengthens when loci come closer).

Clearly, this coarse-grained description misses many of the specific biochemical mechanisms of loop extrusion, but we will see nevertheless a rich variety of observed phenomena appear. In this paper, we will investigate both spatial- and sequence-distance-dependent models but will report primarily on results from the sequence-distance-dependent model in the main text. The corresponding results for the spatial-distance-dependent grappling models are provided in \textcolor{blue}{SMs} and proved to be quite similar overall. 

\section{Simulation of Motorized Models of the Mitotic Chromosome}
We first performed the coarse-grained simulations for a homogeneous chromosome polymer at $20$ $kb$ resolution, which matches the typical resolution of Hi-C experiments \cite{lieberman2009comprehensive}. The simulation is implemented using a hybrid method using the Langevin equation to describe the relaxation and thermal diffusion of the system under the homopolymer potential interrupted by additional shot-noise-like grappling events that are handled via the dynamical Monte Carlo algorithm \cite{gillespie1976general,gillespie2001approximate} (see \textcolor{blue}{SMs} and \textcolor{blue}{Methods} for details). By following Refs. \cite{di2018anomalous} and \cite{contessoto2024energy}, the thermal diffusion constant of the coarse-grained beads $D$ are tuned to the experimental values, $2.97$ $\mu m^2/s$ (\textcolor{blue}{SMs}). The grappling distance is set by $l/\sigma=L_{SMC}/L_\sigma\sim0.005$, where $L_\sigma$ is the diameter of beads that represent $20$ kb chromatin sequences as determined earlier by comparing simulated structures to imaging studies \cite{zhang2016shape,contessoto2024energy}. Force extension experiments on condensins and our previous analysis of Hi-C data from motorized chromosomes suggest that SMC motors are quite susceptible owing to the large length scale of the coiled-coil fragments \cite{cao2024motorized,ganji2018real}. They rapidly stall against a modest applied force. Therefore, we set \(\vartheta = 1\) (for both condensin I \& II). The experimental studies suggest that the action range of condensin II in sequence ($\sim 3$ Mb) during loop extrusion is approximately 6-10 times the sequence interaction range of condensin I ($\sim 400$ kb) during prometaphase \cite{gibcus2018pathway,dey2023structural,thirumalai2024organization}. We have assumed that for condensin I, the grappling probability \( \mathcal{C}^I_{ij} \) scales as \(|i - j|^{-1} \) when \( 3 < |i - j| \leq 20 \), but is zero at larger genomic separations. This cutoff range in our $20$ kb resolution simulations is comparable to the average loop size observed in experiments. Condensin I is responsible for forming small-scale arrays of consecutive loops. For condensin II, we assume \( \mathcal{C}^{II}_{ij} \)  also scales as \(|i - j|^{-1} \) when \( 100 < |i - j| \leq 200 \), with zero probability at larger or smaller separations. This range is comparable to the size of the global helical turn observed in experiments ($\sim 3$ Mb), which results in the overall structure exhibiting approximately 10 turns, as seen in DT40 cells. Condensin II is responsible for large-scale correlations to form the rigid axial scaffold. The cutoff ranges of the motor actions are also close to those extracted from the ideal chromosome model based on information theoretical energy landscape theory \cite{chu2020conformational}. The choice of different cutoff ranges will alter the number of helical turns and affect the ratio of the structure's long and short axes. In the spatial-distance-dependent grappling model, we use the same cutoff ranges for the two types of condensins as in the purely sequence-distance-dependent grappling model. However, the grappling probability within the action range decays as \( \mathcal{C}^\alpha(r_{ij}) \propto r_{ij}^{-1} \). Incorporating this direct dependence on distance ensures that grappling events typically occur only when loci are spatially proximate.

The apparent grappling frequencies of condensin I and II are regulated by the kinetic parameters \(\kappa_I\) and \(\kappa_{II}\). When the system is passive with $\kappa_I=\kappa_{II}=0$, the simulated structure will be a simple random flight chain. We monitored the simulations to see whether they had reached a steady state in the simulation runs. We employ a global structural order parameter \( Q \) that characterizes the similarity between the current structure and a referenced configuration \cite{shoemaker1997structural,plotkin1997statistical}. When a single structure dominates, as in protein folding, \( Q \) is an excellent reaction coordinate for describing conformational changes, this is because it quantifies the similarity of a structure to the known funnel-shaped energy landscape characteristic of protein folding. The value of \( Q \) ranges from $0$ to $1$, with higher values indicating greater structural similarity. A stationary value of \( Q \) will indicate fluctuations near a stable motif, suggesting that the system has reached a steady state of structural organization.

\subsection*{The simulation of the purely sequence-distance-dependent model}
Using the timescale established in Refs. \cite{contessoto2024energy,di2018anomalous}, the simulated trajectories here correspond to several hours of laboratory time. This would, for some cells such as DT 40, exceed the duration of a single cell cycle, so the question of history dependence may be important. When the grappling rates of condensin I and II are sufficiently high (large enough \(\kappa_I\) and \(\kappa_{II}\)), the long-time simulated structures nevertheless exhibit characteristics of mitotic chromosomes. Fig. \ref{fig:1}B shows a representative configuration after reaching a steady state when both types of motors are active ($\kappa_I=5$ and $\kappa_{II}=2.5$). We indicate genomic sequence using bead colors that vary along the sequence. The $\kappa_{I,II}$ values are consistent with the apparent extrusion rates from experiments using real-time imaging techniques, etc \cite{ganji2018real}. For these values of $\kappa_{I,II}$, the frequency of grappling events on a single bead corresponding to a 20 kb DNA sequence in our coarse-grained simulations, is approximately 10-20 events per bead per second. These values correspond to an apparent extrusion rate of 1-2 kbp/s, which is comparable to the experimental values 1-3 kbp/s  \cite{ganji2018real,davidson2019dna,kim2019human,ryu2022condensin,bauer2021cohesin,samejima2024rules}. Experiments show that the ATPase consumption by a single condensin complex is roughly 1-2 ATP/s \cite{ganji2018real,davidson2019dna,kim2019human,ryu2022condensin,wang2017bacillus}. The ATP consumption by condensin in grappling thus is quite low compared to the available ATP in cells. The amount of energy consumed by grappling activity needed to set up the mitotic structure is modest compared to other needs of the cell.

We choose the shown representative structure as the reference. The ensemble average of the pairwise similarity order parameters $\langle Q_s\rangle\approx0.35$ (similarity between all pairwise distances) and $\langle Q_c\rangle\approx0.95$ (similarity between pairwise contact probabilities), indicates that the steady-state structure exhibits significant regularity (Fig. \textcolor{blue}{S8A}). The free energy potential $F(Q_s)$ (analogous to the Parisi-Franz potential in spin glass \cite{franz1995recipes,franz1997phase}) has a single basin near $Q_s\approx0.35$ (Fig. \textcolor{blue}{S8B}). The distribution for the $Q$ values between different loci has also been depicted for reference (Fig. \textcolor{blue}{S8C}).  There is not enough time to reach full equilibrium; therefore, the structure will inherently contain defects. As a comparison, we have examined the relaxation time of the representative structure into a random flight chain upon the removal of both motors. The structure expands into a loose spherical shape within a timescale equivalent to a few minutes in laboratory experiments, although some knots can still be observed. As suggested by Hildebrand et al. \cite{hildebrand2024mitotic}, the exit from mitosis involves regulated topoisomerase II activity to untangle the chromatin chain.

We immediately see that the motorized model of the mitotic chromosome naturally adopts a cylindrical shape, as observed under optical microscopy \cite{marko1995stretching}. The structure exhibits striking rotational symmetry breaking with clear anisotropy. The plotted phase diagram (Fig. \textcolor{blue}{S10}) shows that variations of the ratio of the largest and smallest extension length along the principal axes from 1 and global chiral symmetry breaking emerge when both \(\kappa_I\) and \(\kappa_{II}\) are sufficiently high, but these geometrical measures eventually saturate. This indicates that within a certain range of \(\kappa_I\) and \(\kappa_{II}\), the overall structure indeed will exhibit the major mitotic chromosome structural characteristics, namely, a cylindrical structure with chiral symmetry breaking. In addition to quasi-periodic local helices, the overall structure distinctly shows fiber-like features with long-range correlations, indicating the establishment of cholesteric liquid crystal ordering in the mitotic chromosome model. The layered liquid crystal fiber structure and helical conformations are consistent with observations from optical microscopy experiments \cite{gibcus2018pathway,de1988metaphase,kireeva2004visualization,kubalova2023helical,belmont1987three,sedat2022proposed1,sedat2022proposed2,mcdonald2024helical}.  Such layering can also be observed in direct inversion by information theoretical simulation based on Hi-C data using energy landscape theory \cite{zhang2016shape,contessoto2024energy}. The simulations based on the Hi-C data turn out to be somewhat more disordered than the present results. We should probably take note that the disorder seen in the structure obtained by these inversions may arise during sample preparation or noise in the analysis.  We have also calculated the contact map corresponding to the simulated structure, where the experimentally reported ``second diagonal" clearly appears (Fig. \textcolor{blue}{S7A}). Unlike the earlier studies using the agnostic information theoretical Hamiltonians inferred from the prometaphase Hi-C data  \cite{zhang2016shape,contessoto2024energy}, the model allows us to explicitly assign the mechanism and action of two kinds of motors. Our representative structures resemble the results obtained using the energy landscape model inferred from Hi-C data but display greater regularity, even without employing quenching protocols. 

\subsection*{The simulation of the spatial-distance-dependent grappling model}
We also simulated the spatial-distance-dependent grappling motor model, where the grappling probability explicitly depends on the spatial distance between two sites. When the grappling rates of condensin I and II are sufficiently high (large enough \(\kappa_I\) and \(\kappa_{II}\)), the long-time simulated structures also exhibit the observed characteristics of mitotic chromosomes. The phase diagram (Fig. \textcolor{blue}{S10B}) shows that when the grappling rates of condensin I and II are sufficiently high (large enough \(\kappa_I\) and \(\kappa_{II}\)), the simulation shows a similar cylindrical structure with global chirality and two layers of helices as the sequence-distance-dependent model (Fig. \ref{fig:1}D). The simulated structure with spatial distance dependence however appears more disordered and is more compact than that obtained from the sequence-dependent models, which are closer to the results from the energy landscape model \cite{zhang2016shape,contessoto2024energy}. Our previous theoretical analysis has shown that spatial dependence of grapplings will introduce additional effective attractions compared to the sequence-dependent model \cite{cao2024motorized}. While these attractions further compact the chromosome structure they frustrate the local tendency to form local regular helices, making the system more disordered \cite{luthey1995helix}. The simulated contact map still exhibits a distinct second diagonal band as the sequence-dependent model (Fig. \textcolor{blue}{S7D}). Additionally, a robust third diagonal band is observed, only showing a weak signal in the sequence-distance-dependent model. Such a third diagonal band has recently been detected in experiments for the DT40 cells \cite{samejima2024rules}.

\subsection*{The substructure under motorizations}
The bead model we employ is very coarse-grained. Doubtless, motors can act on still shorter sequence scales. We therefore also simulated the effects of grappling motors on substructures by modeling chains with smaller $\sigma/l$ (Fig. \textcolor{blue}{S9}). The simulation results show the potential emergence of dense, liquid crystal-like substructures with second diagonal bands, suggesting the possibility of multilayer fibers on the sub-kilobase scale. However, this hierarchy will be truncated at a critical scale of around 1 kbp unless motors in addition to condensin are used, such as cohesin or phase-separation motors. This truncation occurs because once the grappling distance approaches the size of the induced structure, large-scale reconstructions exerting large forces must take place. The forces applied exceed the condensin's stalling force, leading to the melting of compacted structures at larger grappling distances. The ATP-utilizing chromatin assembly and remodeling factor (ACF) from the ISWI family is known to drive anti-correlated kicks at the nucleosome scale ($\sim$332 bp) \cite{zhang2011packing,mobius2013toward}, which can be modeled by grappling motors, which can occur if its stalling force is sufficiently large \cite{jiang2019theory,florescu2012kinetic}. Sedat et al. have proposed a hierarchical helical coiled structure for nucleosome chromosome architectures during mitosis, based on Cryo-EM tomography and computational modeling in this size range \cite{sedat2022proposed1,mcdonald2024helical}. 

\subsection*{Quantitative analysis of the mitotic chromosome structure}
To quantitatively study the helical features and chirality of the chromosome, we examined two order parameters of the static structure. First, we calculated the orientation order parameter $O_{OP}$, which measures the degree of helical twist. This was obtained by calculating the correlation between two unit vectors connecting beads \([i, i + 4]\) and \([j, j + 4]\) along the chromosome chain (see \textcolor{blue}{SMs} for details). In Fig. \ref{fig:1}C and E, we present the Fourier transform of the $O_{OP}$ signal in the genomic spatial frequency domain, with peaks corresponding to characteristic frequencies of helical twisting. The high-frequency peak on the right corresponds to local helices, with each turn spanning approximately several hundred kilobases, consistent with the findings of experimental and simulation results \cite{gibcus2018pathway,zhang2016shape,contessoto2024energy}. The lower-frequency peak on the left indicates layered fibers, spanning several megabases, contributing to the overall chirality. These low-frequency peaks are also related to the ``second diagonal" observed in Hi-C maps. As proposed by Kubalova et al., the ``second diagonal" serves as evidence for the helical organization of chromatin loops \cite{kubalova2023helical}. Similar results have been seen in the chromosomes of \textit{Aedes aegypti} mosquito, where the chromosome again takes on a liquid crystalline organization \cite{contessoto2023interphase}. The number of larger turns is around ten, consistent with what is seen in typical experiments \cite{camara2024helical}. In the present model, the number of global turns is primarily influenced by \(\mathcal{C}^{II}_{ij}\), indicating a dependence on the range and processivity of condensin II. The global helical signal intensity observed in our simulation results is comparable to that reported in the quenched ideal chromosome model obtained from the Hi-C data, but the local helical signal is much stronger than what results from the direct information theoretical inversion of the mitotic chromosome. The intensity of the signal has increased by 20-fold \cite{zhang2016shape,contessoto2024energy,zhang2015topology}.

In addition to monitoring helix formation with the orientational order parameter, we introduce a chiral order parameter $\Psi_i$ for the local genomic locus $i$. The details are outlined in \textcolor{blue}{SMs}. In Fig. \ref{fig:1}B and D, we show the variations in local chirality within the structure, indicating that the motorized model of the mitotic chromosome overall exhibits strong global chirality. This is an inevitable result of the layered helical formation. We see however that the chiral order is imperfect. There are interesting half-turns or ``perversions" within the structure. Such perversions have sometimes been seen in light microscopy experiments \cite{chu20203d,chu2020one}. These disinclinations can be thought of as defects within the cholesteric phase, where the chromosome chain undergoes a pronounced reversal in chirality, accompanied by other types of defects, such as tilt and twist grain boundaries \cite{read1950dislocation}. We will further analyze the mechanism and nature of these defects in the following sections.

Our simulations reveal another intriguing finding: the motorized mitotic chromosome structure contains some knots but overall it is not heavily entangled. The topology of chromosomes, particularly the knots, has long been a subject of significant interest \cite{grosberg1993crumpled,grosberg1988role,halverson2014melt,polovnikov2023crumpled,tubiana2024topology,tamm2015anomalous,orlandini2019synergy,racko2018chromatin,conforto2024fluidification} due to its biological relevance noted already by Waston and Crick in their first papers on DNA \cite{watson1953molecular,crick1976linking}. A knot-free chromosomal structure will facilitate replication as well as target-search \cite{kim2022fractal}. Even the simplest cellular organisms have evolved topoisomerases to address the knotting problem \cite{champoux2001dna}. Complex entangled conformations would pose kinetic barriers to DNA replication and transcription. By analyzing the simulated structure's corresponding Alexander polynomial, we estimate that the number of knots is around $20$ to $30$, far lower than what would be predicted for compacted free flight chain polymers \cite{grosberg1996flory}. 

In summary, we have described the mitotic chromosome structures that emerge from the motor activity of both models of condensin I and condensin II, and extensively compared them with experimental observations. In the following sections, we will pull apart the mechanism by separately studying the effects of condensin I and condensin II, demonstrating that both motors are required to replicate experimental results. The roles of condensin I and II during mitosis are quite distinct, despite their shared classification as SMC motors and similar apparent mechanisms of action.

\section{The effect of condensin II during mitosis}
\subsection*{Condensin II-depletion simulations}
\begin{figure*}[htbp]
\renewcommand*{\thefigure}{\arabic{figure}}
\centering
\includegraphics[width=1\columnwidth]{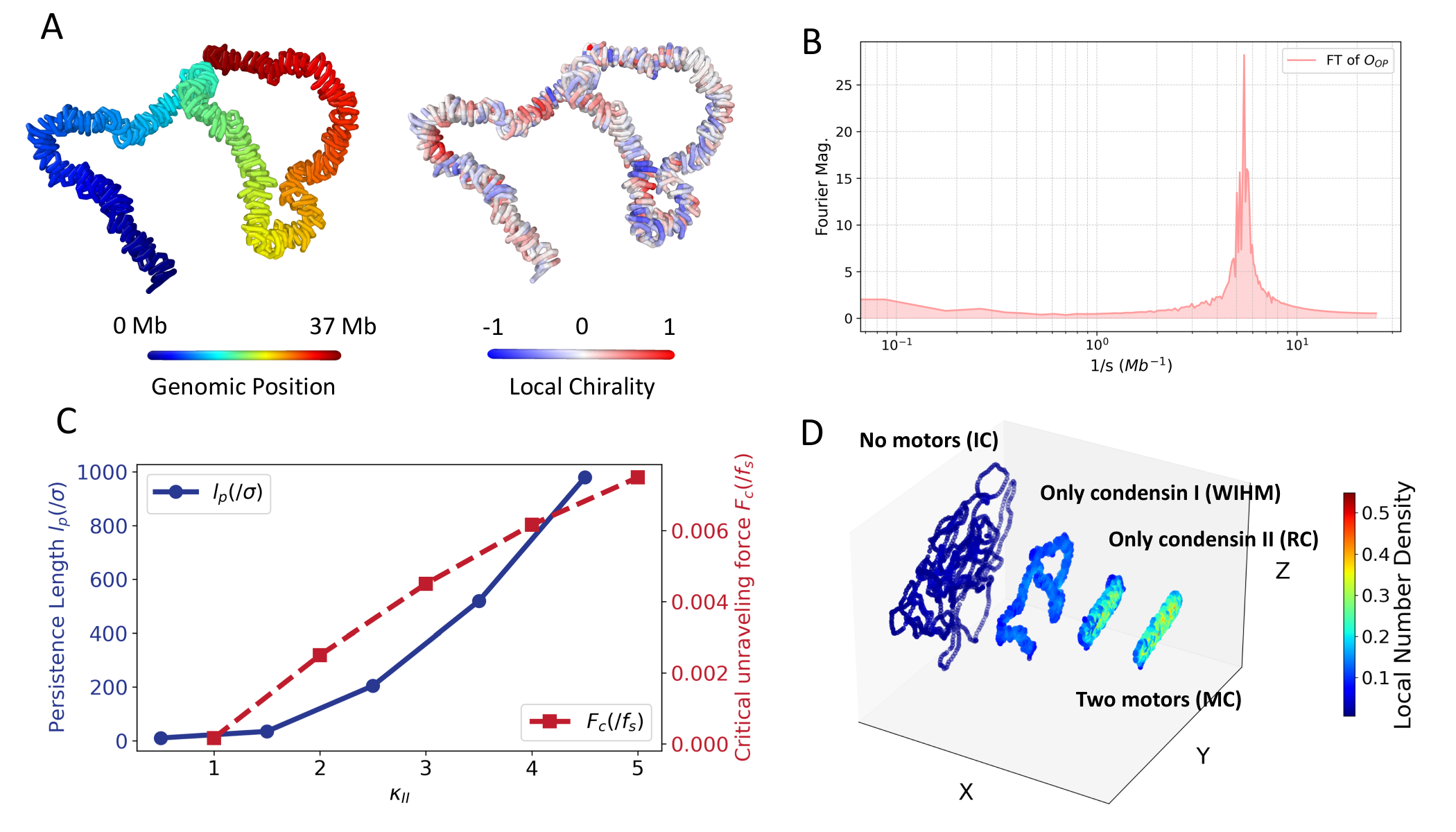}
\caption{\textbf{The effect of condensin II during mitosis.} (A) A representative structure of the condensin II-depletion simulations, where only condensin I is active ($\kappa_I=5$ and $\kappa_{II}=0$). The color bars indicate the color (left panel) and local chirality (right panel) variations along the genomic sequence. (B) The Fourier transform of the representative structure's orientational order parameter along the genomic position. (C) The persistence length $l_p(/\sigma)$ (blue circles and solid line) and the critical unraveling force $F_c$ normalized by the stalling force $f_s$ of grappling motors (red squares and dashed line). Under the action of condensin II, mitotic chromosomes develop a rigid axial scaffold and ensure their structural integrity under stretching forces. (D) The local number density of different representative structures. From left to right, they are interphase chromosomes ($\kappa_I=\kappa_{II}=0$), worm-like ideal helix model ($\kappa_I=5$, $\kappa_{II}=0$), random cylinders ($\kappa_I=0$, $\kappa_{II}=5$) and mitotic chromosomes ($\kappa_I=\kappa_{II}=5$). Condensin II activity significantly compacts the chromosomal conformation.}
\label{fig:2} 
\end{figure*}

We first examine the scenario where only condensin I is active by setting \(\kappa_{II} = 0\). In Fig. \ref{fig:2}A, we show a representative structure where condensin I, with its short-range grappling effect, extrudes orderly local helices. In this case, however, the absence of condensin II leads to a lack of long-range correlations, making the overall configuration appear random. Local helix formation enhances the chromosome chain's local rigidity, preventing it from bending freely like a fully flexible chain (similar to a worm-like chain \cite{kratky1949rontgenuntersuchung}). Therefore, we refer to this structure as a ``worm-like ideal helix model" (WIHM). Overall the random flight nature of the renormalized chain is quite distinguishable from the cylindrical structure usually seen in mitosis.

In Fig. \ref{fig:2}B, for the case with only condensin I, we plot the Fourier transform spectrum of the $O_{OP}$ and the variations in local chirality. The $O_{OP}$ spectrum now only exhibits a high-frequency peak on the right, reflecting small rather local helices with sizes around several hundred kilobases. The local chirality fluctuates around zero, indicating an absence of global chirality, which is also evident from the disappearance of the low-frequency peak in the $O_{OP}$ spectrum. Our results, where condensin II is depleted, suggest that condensin II is essential for the formation of the liquid crystal-like cylinder. This finding is consistent with the observations of Gibcus et al. \cite{gibcus2018pathway}.

\subsection*{Condensin II assists in forming a rigid scaffold in the chromosome}
We further characterized the structural features of chromosomes during their transitions from the worm-like ideal helix model to the mitotic chromosome. First, we computed a coarse-grained persistence length \( L_p \) in order to describe the flexibility of the large-scale principal axis scaffold. The persistence length \( L_p \) is directly related to bending stiffness, satisfying the relationship \( B = k_B T L_p \), and is used to characterize chromosomal bending relaxation and mechanical response \cite{poirier2000reversible,poirier2002bending,houchmandzadeh1999elasticity,marko2008micromechanical,sun2018condensin}. The specific calculation method is detailed in \textcolor{blue}{SMs}. In Fig. \ref{fig:2}C, we plot the rapid increase in persistence length as \(\kappa_{II}\) increases, with values ranging from several micrometers to hundreds of micrometers. Simulations based on the energy landscapes inferred from Hi-C data predict that the persistence length of mitotic chromosomes in DT40 cells is around $20$ micrometers, which would correspond to scenarios with relatively low condensin II activity. Experimental measurements of chromatids assembled from \textit{Xenopus laevis} egg extracts also approximate this value \cite{houchmandzadeh1999elasticity}. Under higher activity, the persistence length could reach millimeter scales, close to the results from metaphase chromosomes extracted from live \textit{Xenopus laevis} or newt cells \cite{poirier2000reversible,poirier2002bending,marko2008micromechanical}. Previous studies have shown that the bending rigidity of mitotic chromosomes is much greater than the bending rigidity of interphase chromosomes (approximately ten times higher), causing mitotic chromosomes to behave as stiff elastic rods \cite{zhang2011loops,ruben2023structural,harju2024loop}. Our simulations of motorized chromosomes here more explicitly demonstrate that this rigidity results from the action of condensin II.

\subsection*{Condensin II is crucial for compaction during mitosis}
Next, we illustrate the crucial role of condensin II in compacting chromosomes.  As described in the loop extrusion mechanism, condensin utilizes a ``grappling and clamping" process, continuously drawing nearby DNA segments closer together so that the chromosomes become progressively folded and compacted. We use the local number density $\rho_{local,i}$ (see \textcolor{blue}{SMs} for details) of different chromosomal structures to characterize the degree of compaction (Fig. \ref{fig:2}D). While increases in the grappling rates of condensin I \& II both lead to higher local number density, the effect of condensin II is significantly more pronounced. This suggests that the long-range conformational reorganization induced by condensin II is essential for further compaction during mitosis.

\subsection*{Condensin II helps chromosomes preserve their structural integrity}
We next examined the impact of condensin II on the mechanical response of chromosomes to being pulled macroscopically. During mitosis, chromosomes are stretched and manipulated by spindle fibers, and adequate tensile strength is crucial for maintaining chromosomal integrity. Recent experiments have investigated the mechanical response of single chromatin regions under external force \cite{keizer2022live}. Results indicate that mitotic chromosomes extracted from live cells are highly extensible objects, capable of retaining their shape and intrinsic elasticity even when stretched to five times their original length \cite{strickfaden2020condensed}.

In our simulations, an external force is applied to one end of the cylindrical mitotic chromosome structure, parallel to the main axis of the backbone (see \textcolor{blue}{SMs} for details). At lower stretching forces, the chromosome follows the applied force, as observed experimentally. When the stretching force reaches a critical value, the chromosome unravels. The transition occurs upon the chromosome being stretched to nearly twice its original length, losing structural integrity and displaying a separation into clumps connected by single chromatin fibers \cite{liu2022chromatin}. 

We attribute the maintenance of structural integrity primarily to the long-range grappling motions mediated by condensin II. To illustrate this, in Fig. \ref{fig:2}C, we plot the variation of critical force with \(\kappa_{II}\), showing a positive correlation between the two. Specifically, the chromosome can better maintain structural integrity as condensin II activity increases. Referring to the analysis in Ref. \cite{ruben2023structural}, the critical force is approximately $0.087$ nN, slightly below the force required in vitro to stretch mitotic chromosomes to twice their original length ($0.1$ to $1$ nN) \cite{marko2008micromechanical}. We calculated the frequency of grappling events. We found that this quantity is nearly independent of the magnitude of the stretching force (even beyond the critical value) and is primarily influenced by motor activity (Fig.\textcolor{blue}{S6}).

We can obtain the mitotic chromosome's Poisson ratio, which is a material property that describes how the chromosome's shape responds to the pulling. The Poisson ratio is the ratio of the lateral strain to the axial strain when a material is stretched or compressed. Within the elastic theory, this material property determines how much the chain distorts when external stress is exerted. This ratio quantifies the tendency of a material to contract in directions perpendicular to the applied force when it is stretched or to expand when compressed. The Poisson's ratio in our simulation $\nu$ varies over the range from $0.102$ to $0.216$ as condensin II becomes more active ranging from $\kappa_{II} =2$ to $\kappa_{II} =5$. These simulated values are close to the experimental Poisson ratio $\sim0.1$ \cite{poirier2000reversible,poirier2001probing}. This rather low Poisson ratio is close to glasses ($\sim0.18-0.3$ \cite{nemilov2007structural}), rivaling that of cork \cite{fortes1989poison}.

\subsection*{Simulations including additional attractions}
At the end of this section, we present an important control experiment where an attractive contact potential (possibly due to interactions with phase-separating proteins \cite{goychuk2023polymer,hnisz2017phase,shin2017liquid}) is introduced into the system (Fig. \textcolor{blue}{S4}). The resulting structure displays a compact configuration with a slightly elongated shape, having a long-to-short axis ratio slightly greater than $1$. Further analysis reveals that, while the structure does exhibit local helices and some degree of chirality, the overall configuration does not form an ordered helix. This lack of symmetry breaking is reflected in the near-zero average global chirality and the absence of low-frequency peaks in the $O_{OP}$ spectrum. Additionally, we plotted the contact map of this structure,  which now shows only weak ``second diagonal" signals (Fig. \textcolor{blue}{S7F}). These findings further support there being critical roles of condensin II.

\section{The effect of condensin I}
\subsection*{Condensin I-depletion simulations}
\begin{figure*}[htbp]
\renewcommand*{\thefigure}{\arabic{figure}}
\centering
\includegraphics[width=1\columnwidth]{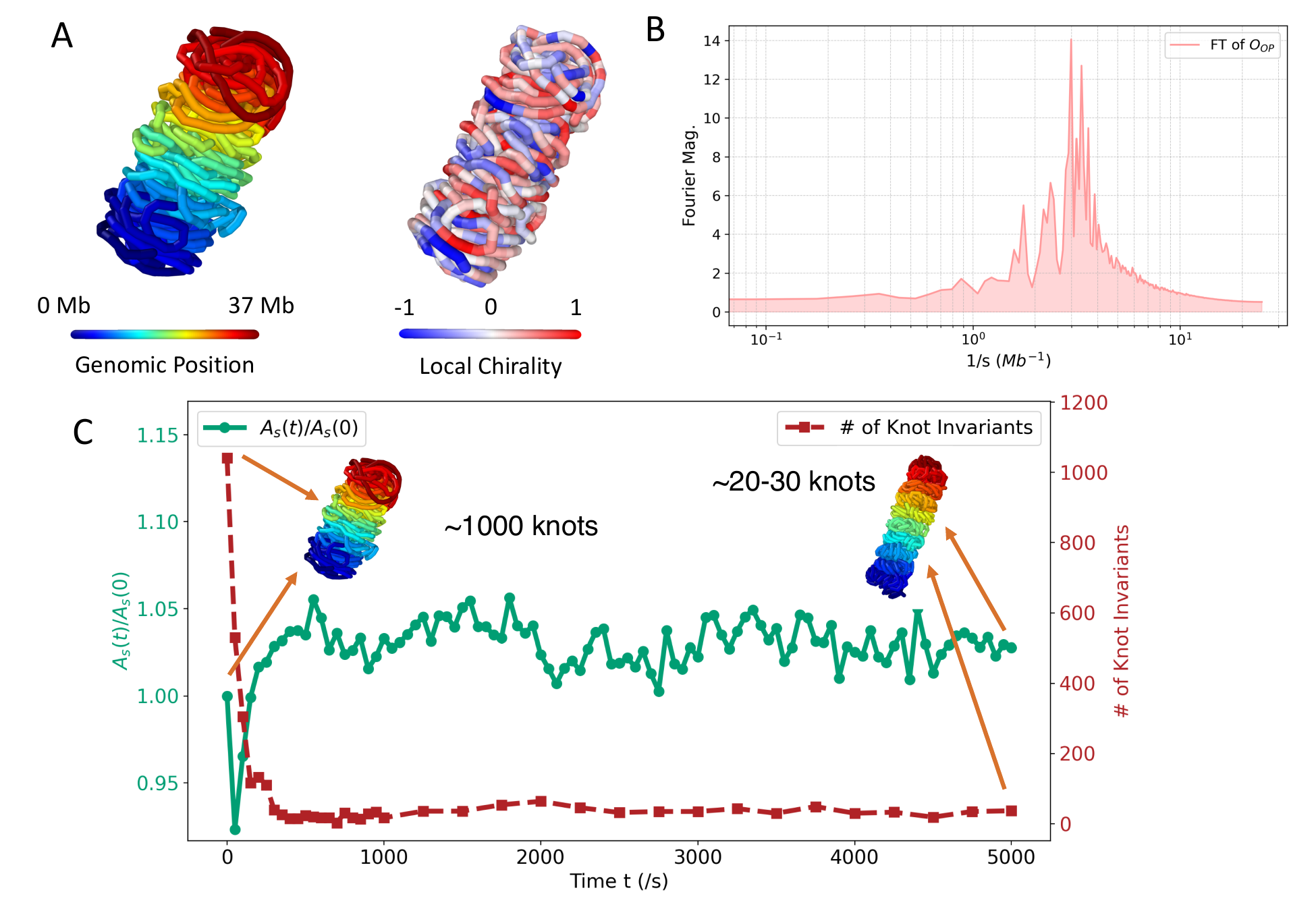}
\caption{\textbf{The effect of condensin I during mitosis.} (A) A representative structure of the condensin II-depletion simulations, where only condensin II is active ($\kappa_I=0$ and $\kappa_II=5$). The color bars indicate the color (left panel) and local chirality (right panel) variations along the genomic sequence. A weak, short-range attraction is included here. (B) The Fourier transform of the representative structure's orientational order parameter along the genomic position. (C) Relaxation of the random cylinder's topology is measured with the number of knot invariants and the apparent accessible surface area $A_s(t)$ normalized by the initial value $A_s(0)$ as a function of time. Here, the condensin I is opened at $t=0$, and the topoisomerases are sufficiently active. (Insets) The initial and final chromosomal conformations. With the assistance of topoisomerases, condensin I can partially untangle chromatin chains, thereby enhancing their accessibility.}
\label{fig:3} 
\end{figure*}

In this section, we investigate what would happen only condensin II were to be present. We first consider the scenario with additional contact interactions. The results without them will be presented later. Fig. \ref{fig:3}A shows a representative structure, displaying a cylindrical shape. Compared to the mitotic chromosome structure obtained from the full motorized model shown in Fig. \ref{fig:1}, the local helices in this configuration have become blurred. This blurring can be quantified by the $O_{OP}$ spectrum in Fig. \ref{fig:3}B, where the peak position remains close to the high-frequency peak in Fig. \ref{fig:1} but is slightly shifted toward lower frequencies, however with considerably increased peak width. This indicates that the local helices are larger and irregular. In Fig. \ref{fig:3}A, we plot the variations in local chirality, which now show chaotic and intense oscillations, suggesting an absence of global chirality in the structure. We referred to this structure as a ``random cylinder" (RC). Additionally, we present the contact map of the structure, which displays a weak ``second diagonal" signal, indicating that the structure remains significantly heterogeneous (Fig. \textcolor{blue}{S7}).

We note that the formation of this structure arises from the competition between condensin II-induced compaction and local attraction, where the local attraction tends to form short-range helices \cite{chan1990origins}. Much as in the molten globules of helical proteins \cite{luthey1995helix}, where the spatial constraints caused by packing in proteins would lead to inevitable frustration in the formation of helical order in homopolymers. This non-discriminate compaction by contact interactions frustrates the partially helical order induced by condensin II. 

\subsection*{Condensin I disentangles knots with the assistance of topoisomerase}
We note that a key difference between the configurations generated without condensin I and those generated using both condensins is the presence of a large number of knots that would form in the absence of condensin I, approximately $1,000$. Even with topoisomerase leading to only a small barrier for chain crossing, the system is now unable to avoid entanglements during rapid collapse. To illustrate the crucial role of condensin I in the disentanglement of chains, starting from this highly entangled configuration, we activated condensin I at time zero. Fig. \ref{fig:3}C shows the relaxation of the knot invariants in the structure. Our simulations have weak excluded volume interactions, and the presence of a soft-core potential allows chain crossings to mimic the effects of topoisomerases. Using the timescale from Ref. \cite{di2018anomalous}, we find that under conditions with sufficient topoisomerase activity (low barrier), the highly entangled random cylinder resolves its knots within a simulation time corresponding to the prometaphase duration observed in DT40 cells, approximately $8$ minutes in laboratory conditions. When the topoisomerase barrier is increased, however as would happen in the topoisomerase depletion experiment \cite{nielsen2020topoisomerase}, the system only partially disentangles within the cell cycle time, resulting in final configurations having about $200$ to $300$ knots. The details have been outlined in \textcolor{blue}{SMs}. Our results suggest that condensin I is critical for maintaining an unknotted metaphase chromosome. We see that, with the assistance of topoisomerase, condensin I untangles knots on the chromosome, naturally forming the continuous arrangement of local helicity reported by Gibcus et al. \cite{gibcus2018pathway}. Our conclusions align with the experimental observations \cite{baxter2012model} that suggest topoisomerase II alleviates supercoiling stress by cutting and untangling double-stranded DNA, and interacts with the condensin complex, facilitating chromosomal remodeling and segregation.

\subsection*{Condensin I enhances chromatin accessibility}
The assistance of condensin I in untangling facilitates chromatin relaxation, making the chromatin structure more open and thereby increasing chromatin accessibility. To quantitatively illustrate this, Fig. 3C shows the changes in the apparent accessible surface area $A_s$ of the chromosome during the untangling process described in the previous paragraph. We observe that $A_s$ initially decreases right after condensin I is activated, due to motor-driven compaction. Subsequently, as knots become untangled, $A_s$ increases rapidly. The final steady-state configuration has a greater $A_s$ than the initial random cylindrical configuration. Our results suggest that the synergistic actions of condensin I and topoisomerase maintain an open chromatin state, which helps facilitate the binding of transcriptional regulators \cite{nitiss2009dna,joshi2012topoisomerase,pedersen2010histone,pommier2016roles}.

\subsection*{Simulations without local attractions}
As a comparison, we simulated the systems without attractive contact interactions, where the resulting configuration can still be viewed as a random cylinder (Fig. \textcolor{blue}{S3}). The peak width in the $O_{OP}$ spectrum is narrower compared to Fig. \ref{fig:3}B, as the absence of the attractive potential reduces the competitive frustration in the motor-driven loop extrusion process. The results for the chirality order parameter indicate that the structure however still lacks global chiral symmetry breaking.

\section{Structure of the mitotic chromosome}
\subsection*{The hierarchical fiber structure}
\begin{figure}[htbp]
\renewcommand*{\thefigure}{\arabic{figure}}
\centering
\includegraphics[width=1\columnwidth]{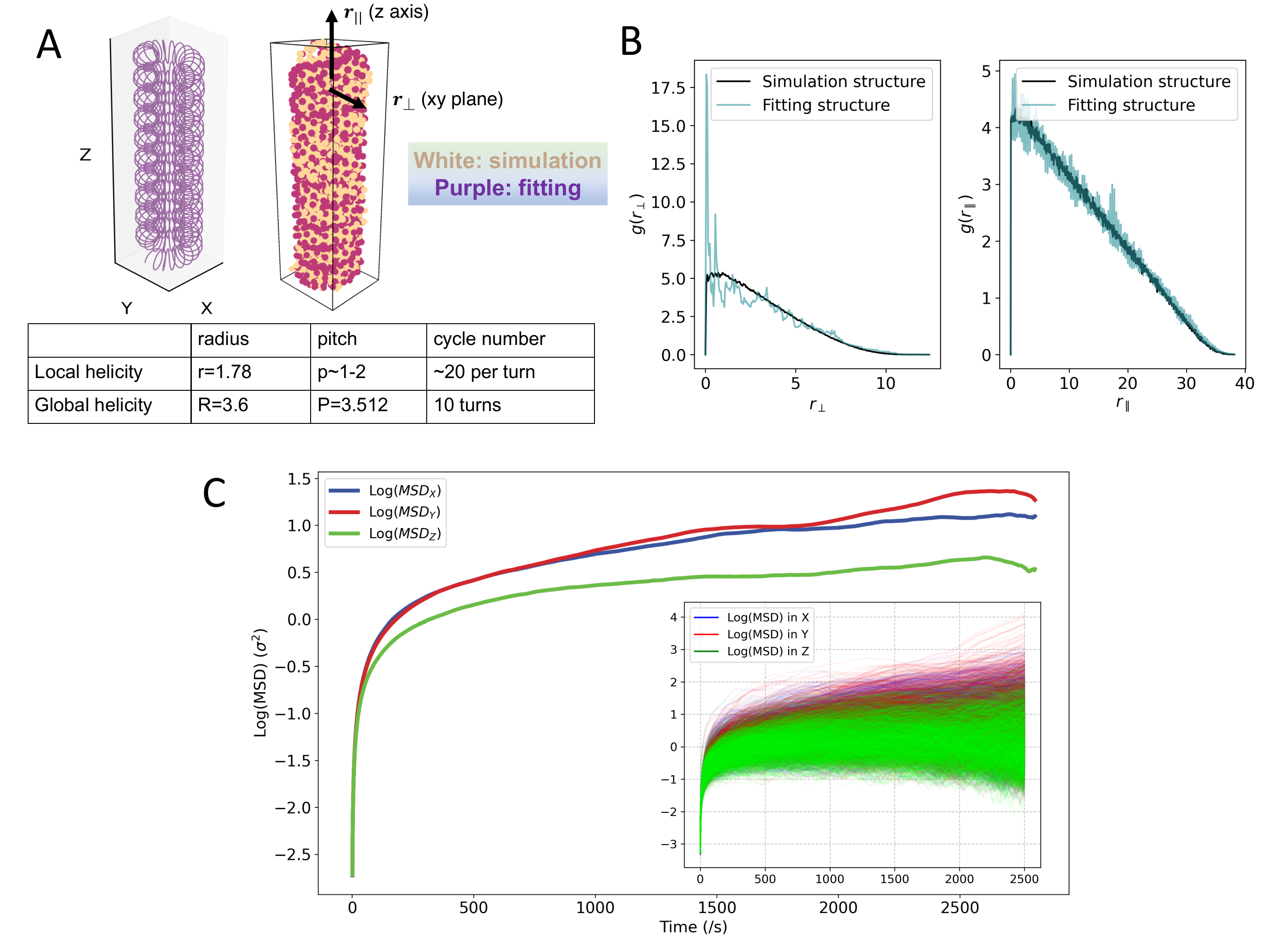}
\caption{\textbf{The organized structure of the mitotic chromosome.} (A) The fitted hierarchical fiber structure (left panel) and its comparison with the simulated mitotic structure (right panel). The fitting structure is shown in purple and the simulated structure is shown in white. (Table) The parameters are extracted from the fitting by using the hierarchical fiber as a referenced structure. (B) The pair distribution functions \( g(r_\perp) \) and \( g(r_\parallel) \) for both the simulated and fitted structures. (C) The averaged mean squared displacement (MSD) of particles in the structure along the \(x\), \(y\), and \(z\) directions. The MSDs in the \(x\) and \(y\) directions, \( \text{MSD}_{x,y} \), quantify vibrations within the primary plane, while the \(z\)-direction MSD quantifies fluctuations along the main axis. The MSDs are averaged over beads and reach plateau values for large $t$. (Insets) The squared displacements of the single locus.}
\label{fig:4} 
\end{figure}

The mitotic chromosome structure shown in Fig. \ref{fig:1} can be considered as a ``fiber of fibers", exhibiting both fractal and liquid crystal phase characteristics. The idea that mitotic chromosomes exhibit a hierarchical helical structure can be traced back to Crick and coworkers \cite{bak1977higher}. If sufficiently specific, we could say the structure is crystalline. Here we used a single hierarchical fiber model to fit this structure, with fitting parameters including the global helix radius \( R_g \), pitch \( P_g \), local helix radius \( r_l \), and the turn numbers of the local and global helices, \( n_l \) and \( N_g \) (see Fig. \ref{fig:4}A and \textcolor{blue}{Methods}). 

In Fig. \ref{fig:4}A, we compare the simulated structure and the fitted structure, along with the resulting values of the fitting structural parameters. The ratio of the global helix radius to the local helix radius is approximately $2.0$, slightly smaller than the ratio obtained from the information theoretical energy landscape simulations based on Hi-C data-driven interactions (around $2.8$) \cite{zhang2015topology}. 

The hierarchical fiber structure can be described in cylindrical coordinates \((r_\perp, \theta, r_\parallel)\), where \( r_\perp \) is the radial distance in the plane perpendicular to the main axis (the primary plane), \( \theta \) is the rotational angle within this plane, and \( r_\parallel = z \) represents the main axis coordinate. In a perfect hierarchical fiber, isotropy is maintained along the \(\theta\)-direction, and we can use the pair distribution functions \( g(r_\perp) \) and \( g(r_\parallel) \) to characterize the structure in the primary plane and along the main axis. In Fig. \ref{fig:4}B, we plot \( g(r_\perp) \) and \( g(r_\parallel) \) for both the simulated and fitted structures. We observe that their pair distribution function curves almost collapse, indicating that the mitotic chromosome structure has a hierarchically layered organization. The local peaks in the pair distribution function, which would appear in a perfect hierarchical fiber, become smaller and more diffuse in the simulated structure, as fluctuations blur the local structures.

\subsection*{The mean squared displacements}
Next, we quantified deviations from the perfect crystal structure using the averaged mean squared displacement (MSD) of particles in the structure along the \(x\), \(y\), and \(z\) directions (ensuring that the main axis of the structure is aligned perpendicular to the \(z\)-axis). The MSDs in the \(x\) and \(y\) directions, \( \text{MSD}_{x,y} \), quantify vibrations within the primary plane, while the \(z\)-direction MSD quantifies fluctuations along the main axis. The MSDs, which are averaged over beads, reach plateau values, with \(x\)- and \(y\)-direction vibrations being nearly identical, supporting isotropy along the \(\theta\)-direction. The \(z\)-direction fluctuations are weaker than particle motions within the layer (Fig. \ref{fig:4}C). We also plotted the squared displacements of the single locus, all of which reach a plateau over long timescales (Fig. \ref{fig:4}C, inset). The plateau values of the single locus exhibit a broad distribution, reflecting the dynamic heterogeneity caused by the structural anisotropy.

We computed the effective spring constants \(\alpha_{\parallel}\) and \(\alpha_{\perp}\) at fiducial locations in a perfect hierarchical fiber structure (\textcolor{blue}{Methods}). Using the plateau behavior of the averaged MSDs at long times, we calculated the Lindemann ratio as \(\alpha_{\parallel}^{-1/2}/a\approx 0.32\) and \(\alpha_{\perp}^{-1/2}/a\approx0.25\), where $a$ is the mean lattice spacing. Both \(\alpha_{\parallel}\) and \(\alpha_{\perp}\) are non-zero positive values, but they do not satisfy the usual Lindemann ratio criterion of around $0.1$. Nevertheless, this structure with liquid crystalline characteristics can still be considered as a rather soft crystal, whose dynamics are vibrations around a single reference state \cite{stillinger1982hidden,debenedetti2001supercooled,singh1985hard}.

\section{Defects in structure}
In the structures we have obtained, the most easily seen prominent defect is the so-called ``perversion". A ``perversion" is a region where the chirality of the helical structure changes from one handedness to another. Such defects may arise from external conditions, such as changes in temperature, pressure, or the application of external fields, or be relics of the formation process \cite{chaikin1995principles}. Defects can also result from local variations mechanical heterogeneity \cite{huang2012spontaneous,liu2016emergent}, material nonlinearity \cite{audoly2000elasticity}, geometric asymmetry \cite{landau2012theory}, or intrinsic curvature \cite{goriely1998spontaneous}.

Perversions belong to a broader class of fundamental defects in systems with discrete symmetry and can be analyzed in relation to other defects in symmetry-breaking structures:

(i). Dislocation: Dislocations are line defects in crystalline structures caused by extra or missing atomic planes. In liquid crystals, dislocations cause local misalignment in molecular arrangements. Perversion regions must contain some dislocations, as changes in chirality require spatial rearrangement of molecules.

(ii). Disinclination: This defect is unique to liquid crystals and involves angular changes in molecular orientation without the addition or removal of material.   Perversions can be viewed as a special type of disinclination, where continuous molecular reorientation leads to a reversal in chirality.


In our simulations, we have noticed that perversions can be resolved through the combined action of the two condensins (see \textcolor{blue}{Movie}), distinguishing these perversions from true topological solitons. During the simulations, the perversions that initially form at the ends of the chain are readily resolved. In contrast, perversions found in the middle of the chain emerge during the formation of the overall ordered structure and exhibit longer lifetimes. Furthermore, we analyzed whether there is a correlation between the locations of perversions and knots. We measured the distance between knots and their nearest perversion across numerous configurations.  The distribution shows no prominent peaks at short distances, indicating a weak correlation between their positions (see Fig. \textcolor{blue}{S5}).

\subsection*{The effects of heterogeneous motorization}
The distinct compartmental characteristics of chromosomes in their initial interphase naturally lead to spatial heterogeneity in condensin activity \cite{nuebler2018chromatin,haarhuis2017cohesin,fudenberg2017fish}. Experimental observations show that condensins are more active in B compartments when mitosis commences \cite{nagano2017cell}. We investigated the impact of heterogeneous motorization on chromosome structure, particularly on the formation of perversions.

\begin{figure}[htbp]
\renewcommand*{\thefigure}{\arabic{figure}}
\centering
\includegraphics[width=0.8\columnwidth]{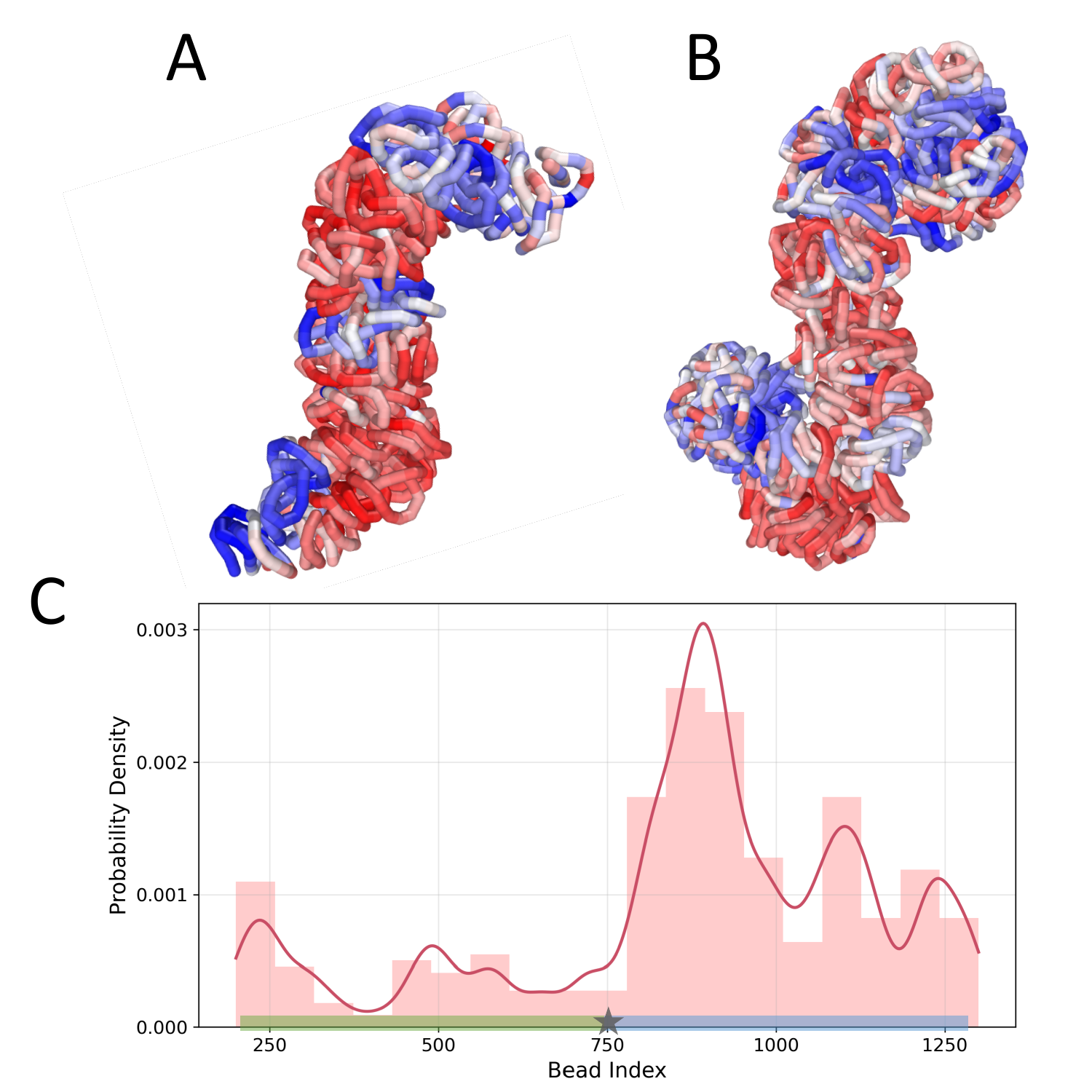}
\caption{\textbf{The effect of heterogeneous motorizations on the perversions.} (A), (B) The representative structure for the heterogeneous motorized model. The heterogeneous motorized model consists of multiple compartments. We assume that motor activity is uniform within each compartment, while the condensin II activity differs between adjacent compartments separated by a junction. (A): 5 compartments, 4 junctions; (B): 10 compartments, 9 junctions. (C) The position distribution of perversions in a diblock copolymer model with heterogeneous motorization. The probability density of perversion positions is displayed in histogram and curve forms. The junction is located at the midpoint of the chain (grey star), with the left segment corresponding to a high condensin II activity region (green rectangle) and the right segment corresponding to a low condensin II activity region (blue rectangle). Condensin I activity is uniformly distributed throughout the chain. The perversion position distribution exhibits a peak near the junction, specifically on the side corresponding to the low-activity region.}
\label{fig:5} 
\end{figure}

We first simulated a heterogeneous motorized chain with high and low condensin II activity, which can be viewed as a type of diblock copolymer. We found that perversions more often form at the junctions between different compartments, typically on the low-activity side (Fig, \ref{fig:5}C). Observing the perversion formation process is reminiscent of the mechanism for knot formation in flexible-rigid diblock polymer ring models, where knots tend to form at the interface between rigid and flexible segments and are compressed within the flexible segment \cite{orlandini2016local}. In our case, the high-activity regions in the chromosome form an overall helical structure much earlier than the low-activity regions and they thus exhibit greater rigidity. In contrast, the low-activity regions, with more flexible chain ends, initiate the formation of the global helix earlier. As a result, the part of the chain near the junction on the low-activity side forms its helical structure last and experiences elastic constraints from both sides during helix formation. The stress constraint between the two regions leads to perversion formation on the low-activity side of the junction. This mechanism also resembles perversion formation in heterogeneous elastic strips \cite{liu2016emergent}.

To explore this idea, we simulated a heterogeneous motorized chain model with multiple compartments (Fig. \ref{fig:5}A,B). The results show an increased occurrence of perversions, all forming near the junctions between compartments. This greater number of perversions seems consistent with observations in mammalian mitotic chromosomes \cite{chu20203d,chu2020one}. We should also note that, as mentioned earlier, the increased number of perversions may also result from attractive contact interactions. As previously found in the information theoretical energy landscape models of the mitotic chromosome, sequence-specific interactions tend to disrupt global chirality \cite{zhang2016shape}.

\section{The pathway for mitotic chromosome formation}
During mitosis, condensin II typically binds during prophase to form the scaffold, followed by condensin I binding in prometaphase to regulate the loop array \cite{gibcus2018pathway}. In Fig. \ref{fig:5}, we illustrate the mitotic chromosome pathway: first, condensin II activity increases, driving the interphase chromosome configuration to form a ``random cylinder" with a rigid scaffold, already exhibiting a rod-like cylindrical structure. This structure is heavily knotted, as observed experimentally \cite{hildebrand2024mitotic,kawamura2010mitotic}. Next, as condensin I activity increases, chromosomes partially untangle, forming a hierarchical fiber-like structure with consecutive loops. Results from the simulations having increased topoisomerase barriers indicate that depleting topoisomerase prolongs the prophase phase, potentially causing cells to exit mitosis without chromosome segregation. This is also consistent with what is seen in the laboratory \cite{nielsen2020topoisomerase}. The simulation results also reasonably predict that excessively high condensin activity leads to a fully collapsed state of the chromosome chain. 

\begin{figure}[htbp]
\renewcommand*{\thefigure}{\arabic{figure}}
\centering
\includegraphics[width=0.8\columnwidth]{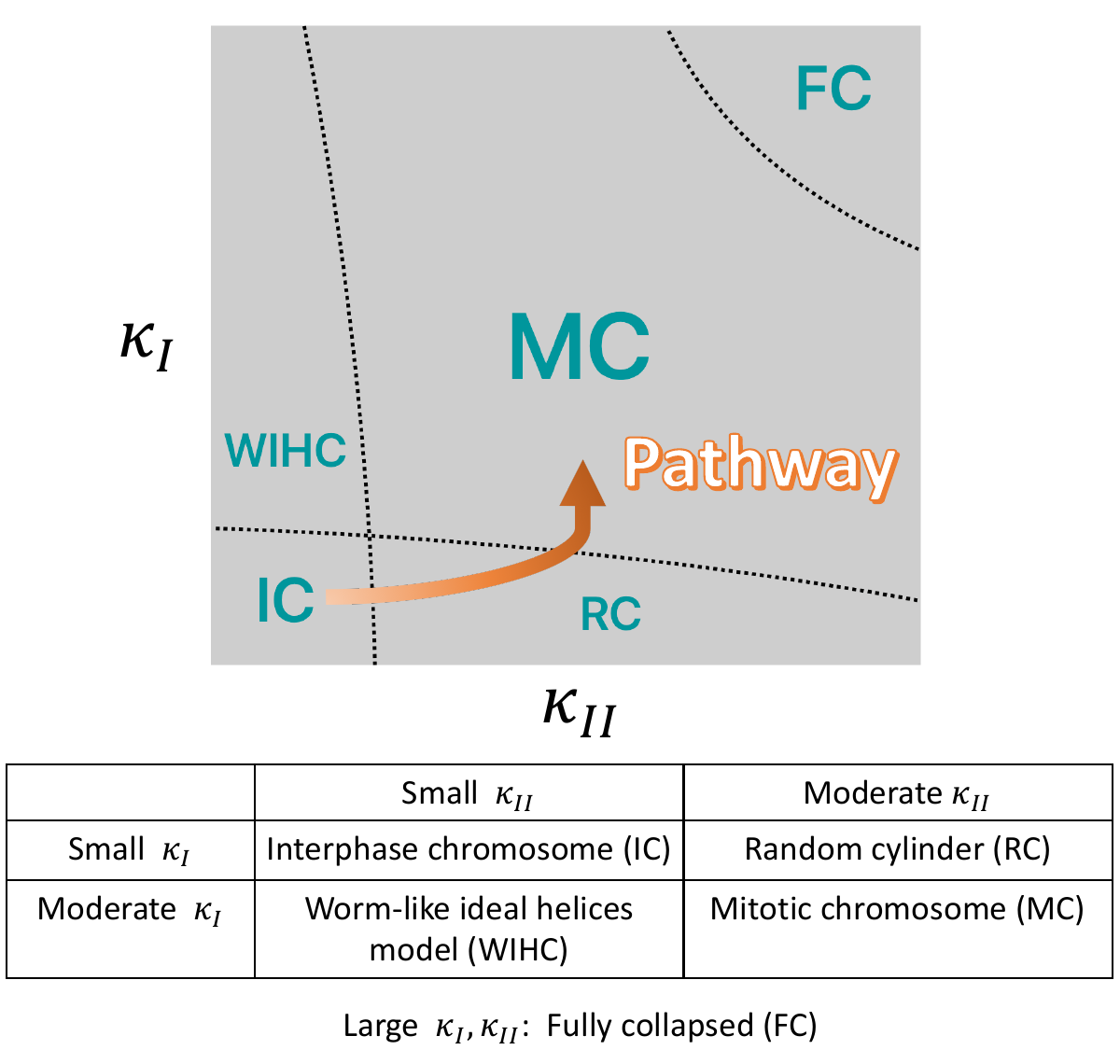}
\caption{The pathway for mitotic chromosome formation. Condensin II binds during prophase, mediating the formation of a highly entangled random cylinder from interphase chromosomes. Condensin I binds during prometaphase and folds the mitotic chromosome in conjunction with condensin II. Excessive condensin activity can lead to the complete collapse of the chromosome chain.}
\label{fig:6} 
\end{figure}

\section{Discussion}
In this paper, we see that an explicit nonequilibrium motor-driven model can reproduce many features of the symmetry-breaking that was seen to occur in quasi-equilibrium models based on an information theoretical energy landscape picture. The structural details correlate with biological knowledge of how distinct motors are introduced at different stages of mitosis and how knots are removed by topoisomerase activity. It will be interesting to perform an information theoretical inversion analysis of the present nonequilibrium simulated structures to see whether the additional ordering that appears here would encode correlations that can be detected in the laboratory. 

The formation of motor-driven mitotic structures is an inherently non-equilibrium process. The condensation-mediated loop extrusion process is ATP-dependent. The grappling motor mechanism we developed operates out of equilibrium, breaking detailed balance and leading to non-zero entropy production \cite{seifert2012stochastic}. The degree of non-equilibrium in the system can be probed through the breakdown of fluctuation-response relations \cite{cugliandolo2011effective,martin2001comparison,mizuno2007nonequilibrium,gnesotto2018broken} and the presence of non-zero curl flux in the order parameter space \cite{gnesotto2018broken,wang2011spontaneous,wang2008potential,battle2016broken}. These explicit nonequilibrium aspects will be explored elsewhere.

\section*{Materials and methods}
\subsection*{Simulation methods}
We carried out hybrid Brownian-Gillespie simulations of the polymer dynamics for the motorized chromosome model: https://github.com/CaOaC/Mitosis. The discrete grappling events described by the Master equation are implemented using the Gillespie algorithm \cite{gillespie1976general}, while the relaxation and thermal diffusion of the chromosome under the homopolymer potential are simulated via the Langevin equation. Additionally, due to the large system size and high number of particles involved in chemical reactions, we use the \(\tau\)-leap method to accelerate the simulation speed of the Gillespie algorithm \cite{gillespie2001approximate}. We simulate for sufficiently long until the contact maps reach stable. Our program allows running on Cuda to use GPU hardware acceleration. Our database includes sample trajectories that can be viewed in OVITO \cite{stukowski2009visualization}. Unless otherwise specified, all simulations start from an interphase chromosome configuration. This is obtained by allowing the structure to relax over time in the absence of motors. This configuration is more relaxed than the typical interphase structure observed in experiments, as cohesin effects are neglected. However, this approach clearly does not affect the applicability of our conclusions. The knot properties of the simulated structures were analyzed using the PyKnot plugin \cite{lua2012pyknot} based on PyMOL \cite{delano2002pymol}, while the apparent accessible surface area was computed using the FreeSASA plugin \cite{mitternacht2016freesasa}.

\subsection*{The hierarchical fiber structure}
Here, we introduce our method to describe the hierarchical fiber structure. We define the radius $R_g$ and pitch $P_g$ of the global helicity. The length of each turn of the global helicity and its angular frequency can be obtained as:
\begin{equation}
    L_g=\sqrt{(2\pi R_g)^2+P_g^2},\quad K_g=2\pi/L_g.
\end{equation}
We then define the radius of the local helicity $r_l$ and the local turns per overall turn $n_l$. The angular frequency and the pitch $p_l$ of the local helicity can be calculated as:
\begin{equation}
    k_l\approx K_gn_l,\quad p_l\approx P_g/n_l.
\end{equation}
The curve for the hierarchical fiber along the contour coordinate in the Cartesian coordinate system can be written as:
\begin{equation}
    \begin{cases}
x(s)=[R_g+r_l\cos(k_ls)]\cos(K_gs) \\
y(s)=[R_g+r_l\cos(k_ls)]\sin(K_gs) \\
z(s)=\frac{P_g}{2\pi}K_gs+r_l\sin(k_ls) , \label{hf}
\end{cases}
\end{equation}
where $s\in[0,s_{max}]$. The parameter $s_{max}$ controls the number of turns of the global chirality $N_g$, which is set to be $10$ in our analysis (DT40 cells \cite{camara2024helical}).

The turn numbers were obtained through statistical averaging of the structure, while other parameters were determined by iteratively minimizing the Kullback-Leibler divergence between the pair distribution function \( g(r_\parallel, r_\perp) \).

\subsection*{The mean squared displacements and effective spring constants}
The diffusion of a single particle is described by the mean-square displacement (MSD). The diffusion in the structure shows anisotropy along the main axis and within the primary plane while being approximately isotropic within the primary plane itself. Thus, we can assume that the steady-state probability distribution of the system can be written as a product of localized Gaussians as
\begin{equation}
    P(\mathbf{r})=\frac{1}{\mathcal{Z}}\prod_i e^{ -\alpha_{\perp}\left[(r_{i,x}-R_{i,x})^{2}+(r_{i,y}-R_{i,y})^{2}\right]-\alpha_{\parallel}(r_{i,z}-R_{i,z})^{2}},
\end{equation}
which often serves as a starting point for the self-consistent phonon analysis \cite{fixman1969highly,stoessel1984linear,singh1985hard}. Here, $\alpha_{\parallel}$ and $\alpha_{\perp}$ are the parallel and perpendicular effective spring constants. $\mathcal{Z}$ is the normalization constant and $\bm{R}_i=(R_{i,x},R_{i,y},R_{i,z})$ represents the fiducial coordinate for the $i$-th bead in the corresponding perfect hierarchical fiber structure.

For a Gaussian distribution, the variance along each axis is inversely related to the localization parameters. In the xy-plane (x and y directions), we have $\sigma_{x}^{2}=\sigma_{y}^{2}=\langle(r_{x}-R_{x})^{2}\rangle=\langle(r_{y}-R_{y})^{2}\rangle={1}/({2\alpha_{\perp}})$. Along the z axis, we have $\sigma_{z}^{2}=\langle(r_{z}-R_{z})^{2}\rangle={1}/({2\alpha_{\parallel}})$. In confined systems, the long-time MSD along each axis will reach a plateau at long times due to the finite variance of particle positions as $\text{MSD}_{j}(\infty)=2\sigma_{j}^{2}$ for $j\in\{x,y,z\}$. Then, we can obtain that
\begin{equation}
    \alpha_{\perp}=\frac{1}{\text{MSD}_{x}(\infty)}=\frac{1}{\text{MSD}_{y}(\infty)},\quad\alpha_{\parallel}=\frac{1}{\text{MSD}_{z}(\infty)}.
\end{equation}

\subsection*{Data, Materials, and Software Availability}
All study data are included in the article and/or SI Appendix. Codes and example output files are available in https://github.com/CaOaC/Mitosis.

\subsection*{Acknowledge}
We would like to thank Vinicius G. Contessoto for the helpful discussions. This work was supported both by the Bullard–Welch Chair at Rice University, grant C-0016, and by the Center for Theoretical Biological Physics sponsored by NSF grant PHY- 2019745.

\subsection*{Author contributions}
Z.C. and P.G.W. designed the research. Z.C., C.D. and Z.H. developed the codes and performed the simulations. Z.C. and P.G.W. contributed to the technical ideas, analyzed data and wrote the paper.

\subsection*{Competing interests}
The authors declare no competing interests.

\bibliography{bibfile}
\end{document}